\documentclass{ws-procs9x6}

\usepackage{epsfig,latexsym}

\begin{document}

\title{Light-hadron electroproduction at next-to-leading order and
implications}

\author{Bernd A. Kniehl}

\address{II. Institut f\"ur Theoretische Physik, Universit\"at Hamburg,\\
Luruper Chaussee 149, 22761 Hamburg, Germany\\
E-mail: kniehl@desy.de}

\maketitle

\abstracts{
We review recent results on the inclusive electroproduction of light hadrons
at next-to-leading order in the parton model of quantum chromodynamics
implemented with fragmentation functions and present updated predictions for
HERA experiments based on the new AKK set.
We also discuss phenomenological implications of these results.}

\maketitle

\section{Introduction}

In the framework of the parton model of quantum chromodynamics (QCD), the
inclusive production of single hadrons is described by means of fragmentation
functions (FFs) $D_a^h(x,\mu)$.
At lowest order (LO), the value of $D_a^h(x,\mu)$ corresponds to the
probability for the parton $a$ produced at short distance $1/\mu$ to form a
jet that includes the hadron $h$ carrying the fraction $x$ of the longitudinal
momentum of $a$.
Analogously, incoming hadrons and resolved photons are represented by
(non-perturbative) parton density functions (PDFs) $F_a^h(x,\mu)$.
Unfortunately, it is not yet possible to calculate the FFs from first
principles, in particular for hadrons with masses smaller than or comparable
to the asymptotic scale parameter $\Lambda$.
However, given their $x$ dependence at some energy scale $\mu$, the evolution
with $\mu$ may be computed perturbatively in QCD using the timelike 
Dokshitzer-Gribov-Lipatov-Altarelli-Parisi (DGLAP) equations.
Moreover, the factorization theorem guarantees that the $D_a^h(x,\mu)$
functions are independent of the process in which they have been determined
and represent a universal property of $h$.
This entitles us to transfer information on how $a$ hadronizes to $h$ in a
well-defined quantitative way from $e^+e^-$ annihilation, where the
measurements are usually most precise, to other kinds of experiments, such as
photo-, lepto-, and hadroproduction.
Recently, FFs for light charged hadrons with complete quark flavour separation
were determined\cite{akk} through a global fit to $e^+e^-$ data from LEP, PEP,
and SLC including for the first time the light-quark tagging probabilities
measured by the OPAL Collaboration at LEP,\cite{opal} thereby improving
previous analyses.\cite{kkp,k}

\begin{figure}[t]
\begin{center}
\includegraphics[width=\textwidth]{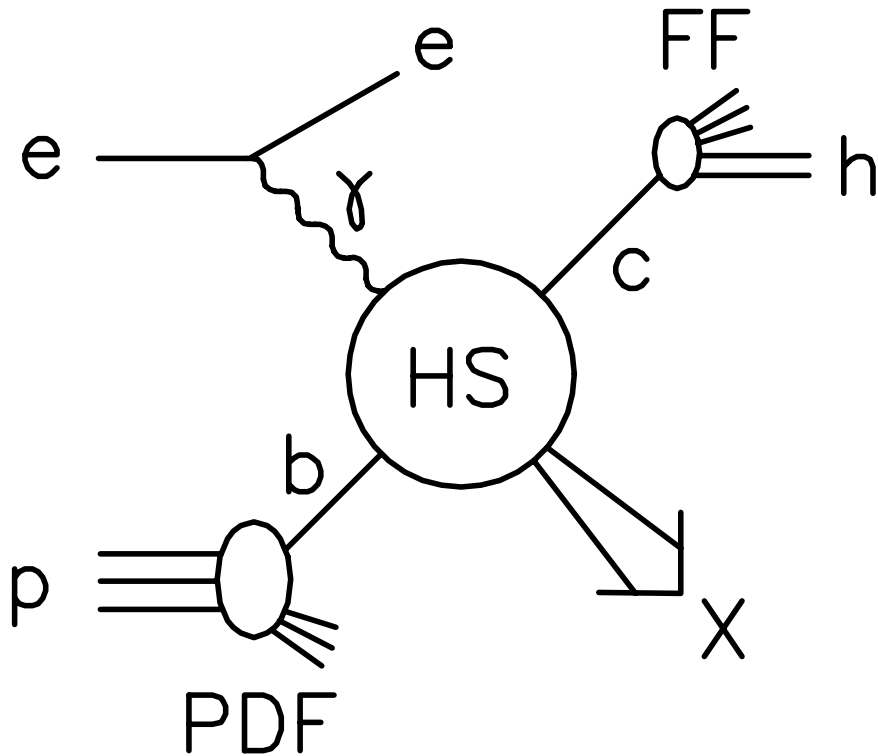} 
\end{center}
\caption{Parton-model representation of $ep\to eh+X$.}
\label{fig:pm}
\end{figure}
The QCD-improved parton model should be particularly well applicable to the
inclusive production of light hadrons carrying large transverse momenta
($p_T$) in deep-inelastic lepton-hadron scattering (DIS) with large photon
virtuality ($Q^2$) due to the presence of two hard mass scales, with
$Q^2,p_T^2\gg\Lambda^2$.
In Fig.~\ref{fig:pm}, this process is represented in the parton-model picture.
The hard-scattering (HS) cross sections, which include colored quarks and/or
gluons in the initial and final states, are computed in perturbative QCD.
They were evaluated at LO more than 25 years ago.\cite{Mendez:zx}
Recently, the next-to-leading-order (NLO) analysis was performed independently
by three groups.\cite{Aurenche:2003by,kkm,Daleo:2004pn}
A comparison between Refs.~\refcite{kkm,Daleo:2004pn} using identical input
yielded agreement within the numerical accuracy.

The cross section of $e^+p\to e^+\pi^0+X$ in DIS was measured in various
distributions with high precision by the H1 Collaboration at HERA in the
forward region, close to the proton remnant.\cite{Adloff:1999zx,Aktas:2004rb}
This measurement reaches down to rather low values of Bjorken's variable
$x_B=Q^2/(2P\!\cdot\!q)$, where $P$ and $q$ are the proton and virtual-photon
four-momenta, respectively, and $Q^2=-q^2$, so that the validity of the
DGLAP evolution might be challenged by Balitsky-Fadin-Kuraev-Lipatov (BFKL)
dynamics.

In Ref.~\refcite{kkm}, the H1 data\cite{Adloff:1999zx,Aktas:2004rb} were
compared with NLO predictions evaluated with the KKP FFs.\cite{kkp}
In Sec.~\ref{sec:two}, we summarize the analytical calculation performed in
Ref.~\refcite{kkm}.
In Sec.~\ref{sec:three}, we present an update of this comparison based on
the new AKK FFs.\cite{akk}
Our conclusions are summarized in Sec.~\ref{sec:four}.

\section{Analytical calculation}
\label{sec:two}

The partonic subprocesses contributing at LO are
\begin{eqnarray}
\gamma^*+q&\to&q+g,\nonumber\\
\gamma^*+q&\to&g+q,\nonumber\\
\gamma^*+g&\to&q+\overline{q},
\label{eq:lo}
\end{eqnarray}
where $q$ represents any of the $n_f$ active quarks or antiquarks and it is
understood that the first of the final-state partons is the one that fragments
into the hadron $h$.

At NLO, processes~(\ref{eq:lo}) receive virtual corrections, and real
corrections arise through the partonic subprocesses  
\begin{eqnarray}
\gamma^*+q&\to&q+g+g,\nonumber\\
\gamma^*+q&\to&g+q+g,\nonumber\\
\gamma^*+g&\to&q+\overline{q}+g,\nonumber\\
\gamma^*+g&\to&g+q+\overline{q},\nonumber\\
\gamma^*+q&\to&q+q+\overline{q},\nonumber\\
\gamma^*+q&\to&\overline{q}+q+q,\nonumber\\
\gamma^*+q&\to&q+q^\prime+\overline{q}^\prime,\nonumber\\
\gamma^*+q&\to&q^\prime+\overline{q}^\prime+q,
\label{eq:nlo}
\end{eqnarray}
where $q^\prime\neq q,\overline{q}$.
The virtual corrections contain infrared (IR) singularities, both of the soft
and/or collinear types, and ultraviolet (UV) ones, which are all regularized
using dimensional regularization with $D=4-2\epsilon$ space-time dimensions
yielding poles in $\epsilon$ in the physical limit $D\to4$.
The latter arise from one-loop diagrams and are removed by renormalizing the
strong-coupling constant and the wave functions of the external partons in the
respective tree-level diagrams, while the former partly cancel in combination
with the real corrections.
The residual IR singularities are absorbed into redefinitions of the PDFs and
FFs.
We extract the IR singularities in the real corrections by performing the
phase space integrations using the dipole subtraction
formalism.\cite{Catani:1996vz}

\section{Comparison with H1 data}
\label{sec:three}

We work in the modified minimal-subtraction ($\overline{\mathrm{MS}}$)
renormalization and factorization scheme with $n_f=5$ massless quark flavors
and identify the renormalization and factorization scales by choosing
$\mu^2=\xi[Q^2+(p_T^\ast)^2]/2$, where the asterisk labels quantities in the
$\gamma^\ast p$ center-of-mass (c.m.) frame and $\xi$ is varied between 1/2
and 2 about the default value 1 to estimate the theoretical uncertainty.
At NLO (LO), we employ set CTEQ6M (CTEQ6L1) of proton
PDFs,\cite{Pumplin:2002vw} the NLO (LO) set of AKK FFs,\cite{akk} and the
two-loop (one-loop) formula for the strong-coupling constant
$\alpha_s^{(n_f)}(\mu)$ with $\Lambda^{(5)}=226$~MeV
(165~MeV).\cite{Pumplin:2002vw}
\begin{figure}[ht]
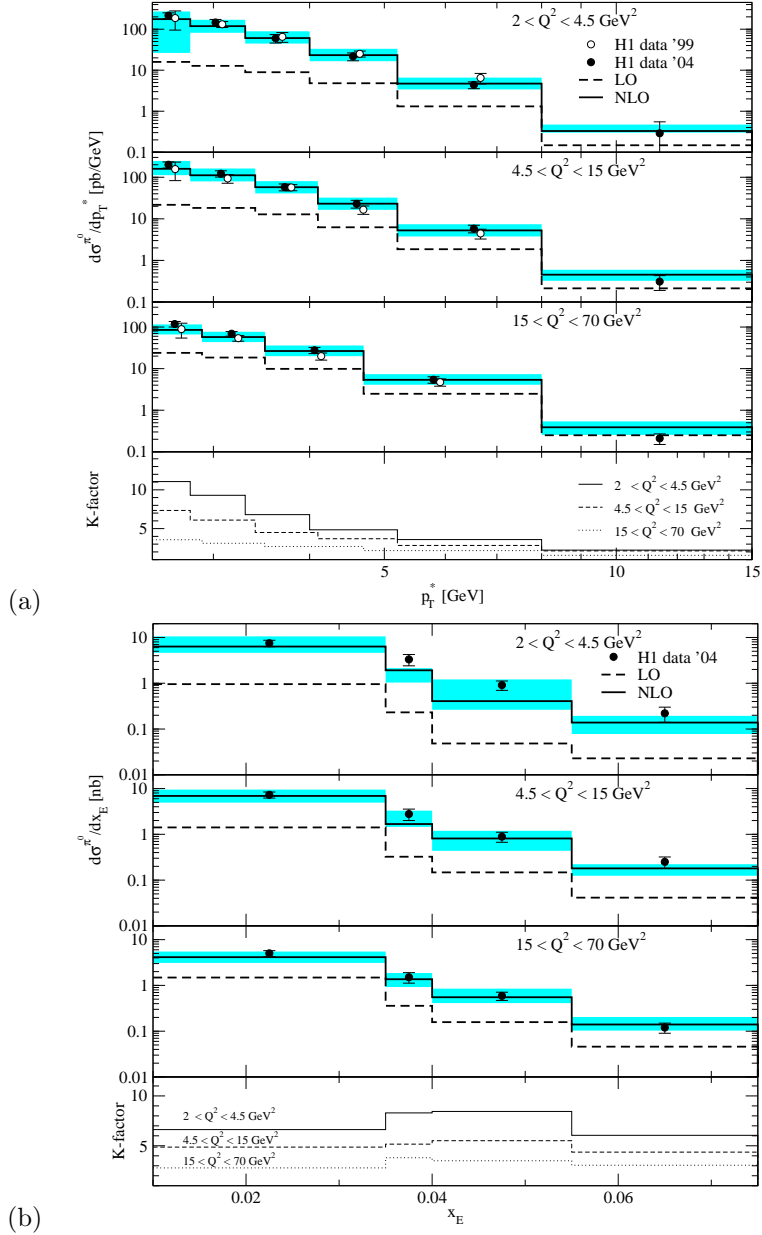

\begin{center}
\begin{tabular}{cc}
(a) & \includegraphics[width=.8\textwidth]{kniehl.fig2.eps} \\
(b) & \includegraphics[width=.8\textwidth]{kniehl.fig3.eps}
\end{tabular}
\end{center}
\caption{H1 data on (a) $d\sigma/dp_T^*$ and (b) $d\sigma/dx_E$ 
for $2<Q^2<4.5$~GeV$^2$, $4.5<Q^2<15$~GeV$^2$, or $15<Q^2<70$~GeV$^2$,
on (c) $d\sigma/dx_B$ for $p_T^*>3.5$~GeV and $2<Q^2<8$~GeV$^2$,
$8<Q^2<20$~GeV$^2$, or $20<Q^2<70$~GeV$^2$, and on (d) $d\sigma/dQ^2$ from
Refs.~\protect\refcite{Adloff:1999zx} (open circles) and
\protect\refcite{Aktas:2004rb} (solid circles) are compared with our default
LO (dashed histograms) and NLO (solid histograms) predictions including
theoretical uncertainties (shaded bands).
The $K$ factors are also shown.}
\label{fig:xs}
\end{figure}

\setcounter{figure}{1}
\begin{figure}[ht]
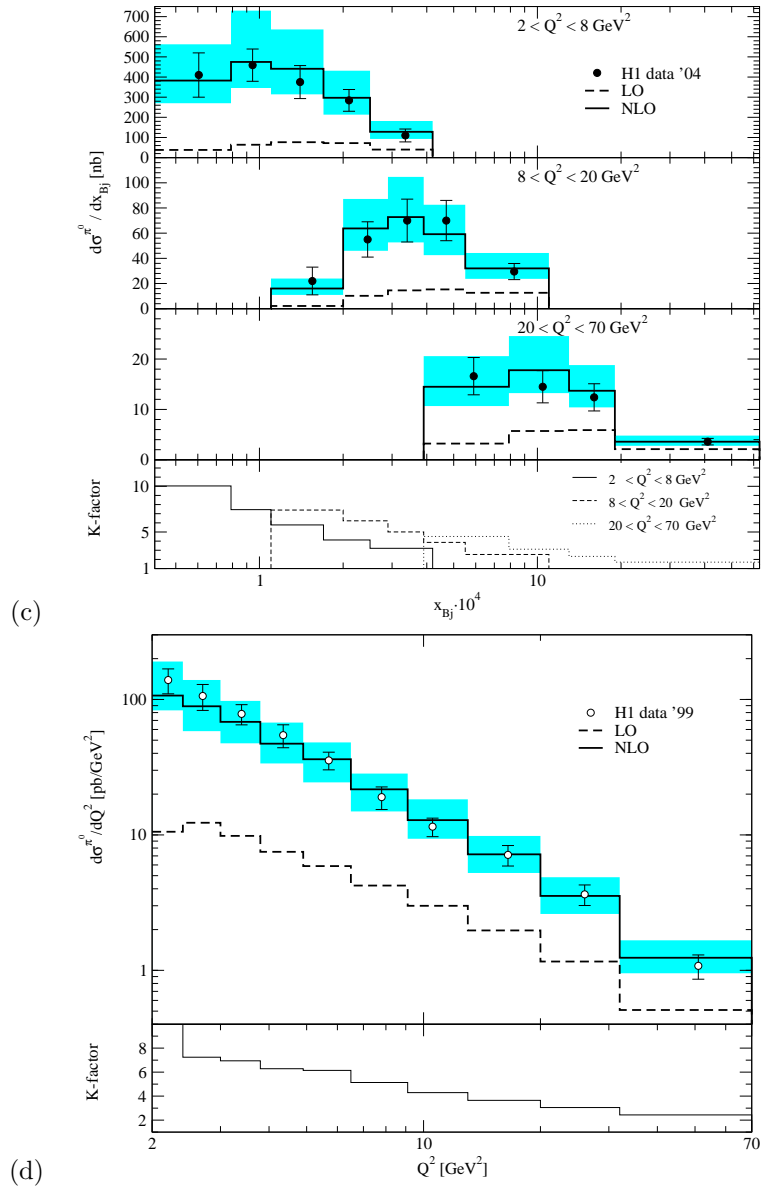

\begin{center}
\begin{tabular}{cc}
(c) & \includegraphics[width=.8\textwidth]{kniehl.fig4.eps} \\
(d) & \includegraphics[width=.8\textwidth]{kniehl.fig5.eps}
\end{tabular}
\end{center}
\caption{Continued.}
\end{figure}
The H1 data\cite{Adloff:1999zx,Aktas:2004rb} were taken in DIS of positrons
with energy $E_e=27.6$~GeV on protons with energy $E_p=820$~GeV in the
laboratory frame, yielding a c.m.\ energy of $\sqrt S=2\sqrt{E_eE_p}=301$~GeV.
The DIS phase space was restricted to $0.1<y<0.6$ and $2<Q^2<70$~GeV$^2$,
where $y=Q^2/(x_BS)$.
The $\pi^0$ mesons were detected within the acceptance cuts $p_T^*>2.5$~GeV
(except where otherwise stated), $5^\circ<\theta<25^\circ$, and $x_E>0.01$,
where $\theta$ is their angle with respect to the proton flight direction and
$E=x_E E_p$ is their energy in the laboratory frame.
The comparisons with our updated LO and NLO predictions are displayed in
Figs.~\ref{fig:xs}(a)--(d).
The QCD correction ($K$) factors, i.e.\ the NLO to LO cross section ratios, are
presented in the downmost frame of each figure.

\begin{figure}[ht]
\begin{center}
\begin{tabular}{cc}
(a) & \includegraphics[width=.675\textwidth]{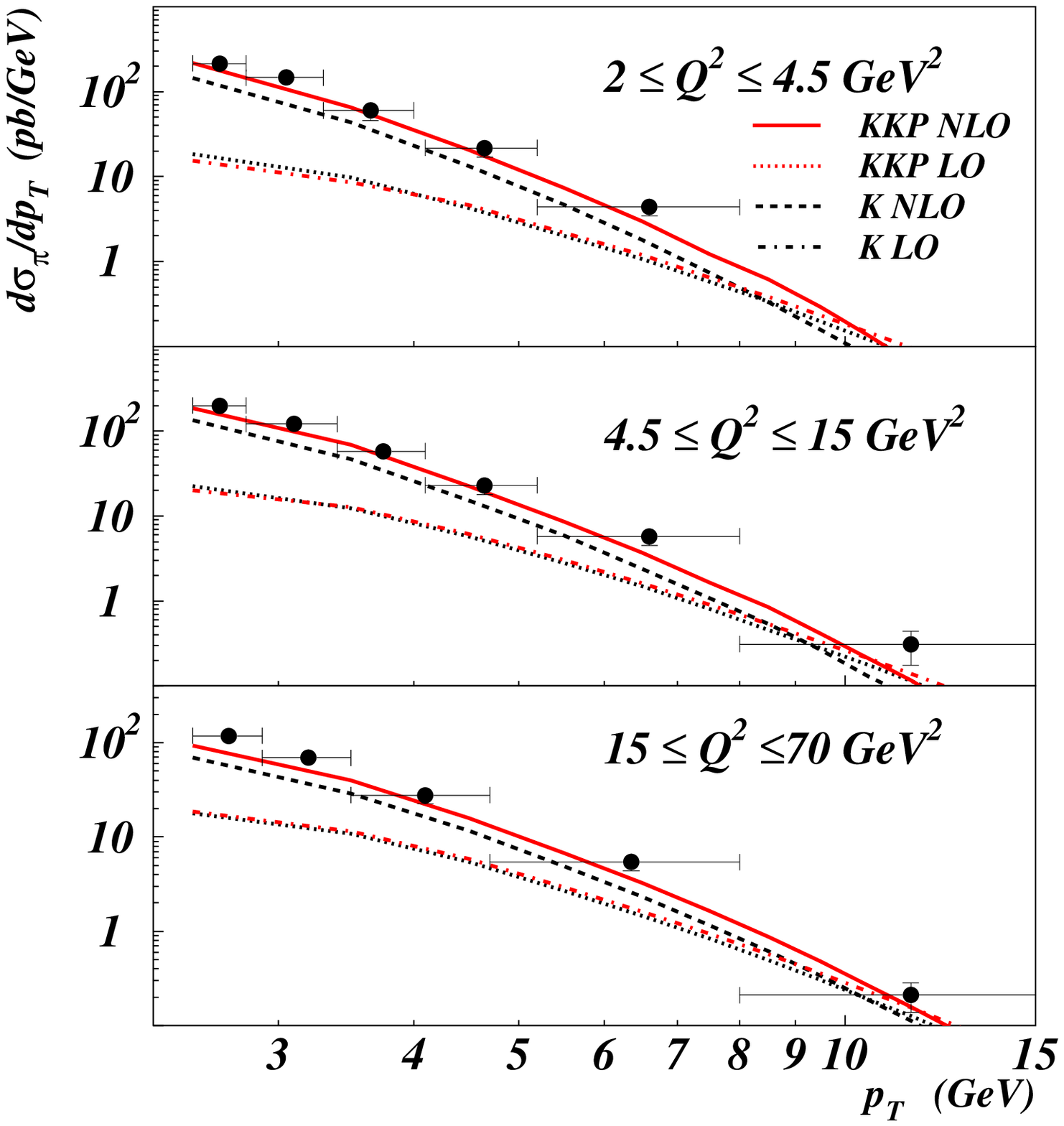} \\
(b) & \includegraphics[width=.675\textwidth]{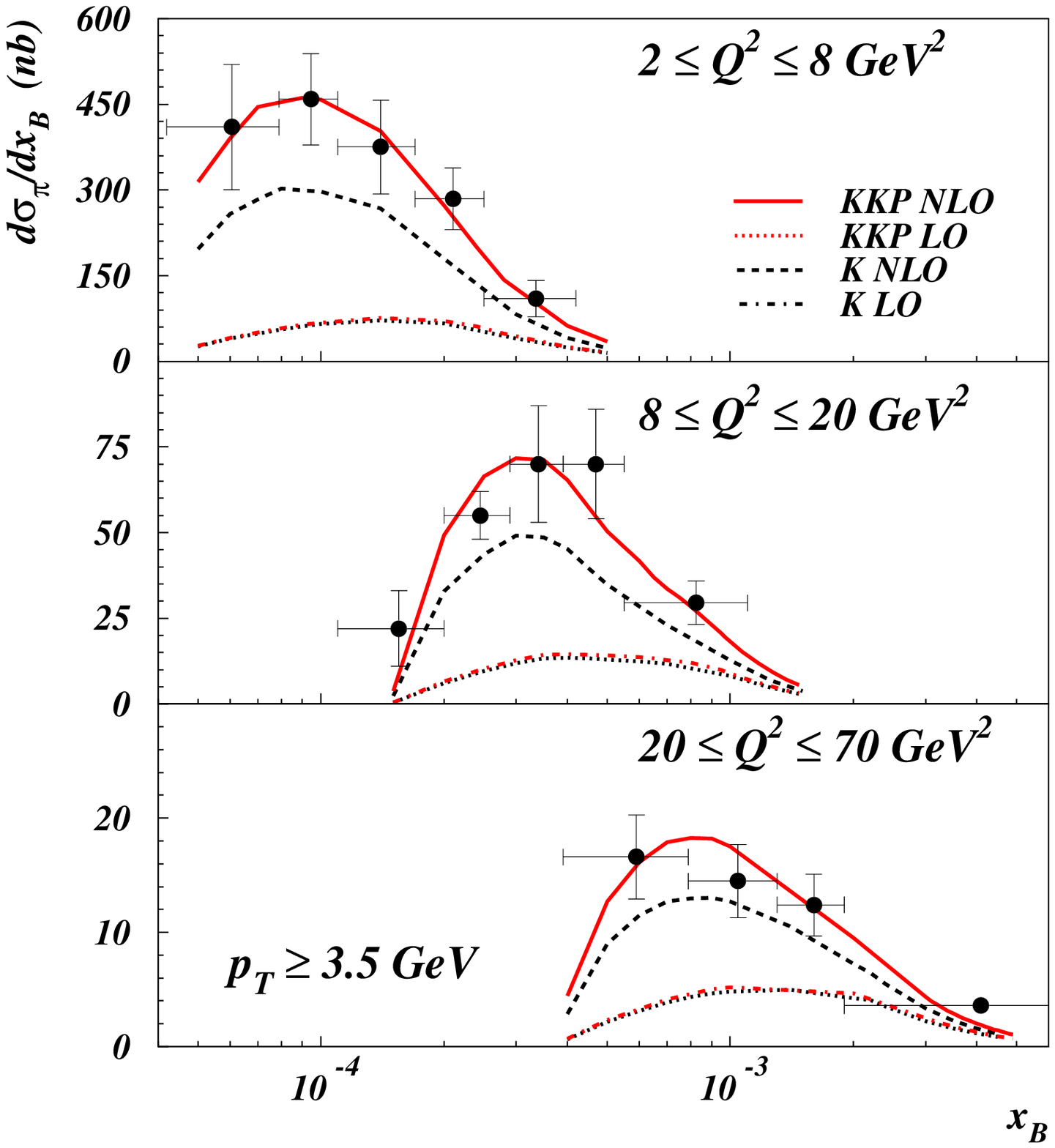}
\end{tabular}
\end{center}
\caption{H1 data\protect\cite{Aktas:2004rb} on (a) $d\sigma/dp_T^*$ for
$2<Q^2<4.5$~GeV$^2$, $4.5<Q^2<15$~GeV$^2$, or $15<Q^2<70$~GeV$^2$ and on (b) 
$d\sigma/dx_B$ for $p_T^*>3.5$~GeV and $2<Q^2<8$~GeV$^2$, $8<Q^2<20$~GeV$^2$,
or $20<Q^2<70$~GeV$^2$ are compared with the LO and NLO predictions evaluated
with the KKP\protect\cite{kkp} or K\protect\cite{k} FFs (taken from
Ref.~\protect\refcite{Daleo:2004pn}).}
\label{fig:ffs}
\end{figure}
Comparison of Figs.~\ref{fig:xs}(a)--(d) with Figs.~3, 5(a), 6(c), and 7 in
Ref.~\refcite{kkm}, where the KKP FFs\cite{kkp} were used, reveals that the
update of our FFs, from set KKP to set AKK,\cite{akk} has hardly any visible
impact on the theoretical predictions considered here.
This may be understood by observing that the OPAL light-quark tagging
probabilities for charged pions,\cite{opal} included in the AKK analysis,
agree well with the assumption made in the KKP one that
$D_u^{\pi^\pm}(x,\mu_0)=D_d^{\pi^\pm}(x,\mu_0)$ at the starting scale $\mu_0$
of the DGLAP evolution.
In Figs.~\ref{fig:ffs}(a) and (b),\cite{Daleo:2004pn} the H1
data\cite{Aktas:2004rb} on $d\sigma/dp_T^*$ for $2<Q^2<4.5$~GeV$^2$,
$4.5<Q^2<15$~GeV$^2$, or $15<Q^2<70$~GeV$^2$ and on $d\sigma/dx_B$ for
$p_T^*>3.5$~GeV and $2<Q^2<8$~GeV$^2$, $8<Q^2<20$~GeV$^2$, or
$20<Q^2<70$~GeV$^2$, respectively, are compared with the LO and NLO
predictions evaluated with the KKP FFs\cite{kkp} or those by Kretzer
(K).\cite{k}
While the LO predictions based on the KKP and K sets agree very well, the NLO
predictions based on the K set appreciably undershoot those based on the KKP
set.
If it were not for the theoretical uncertainty, one might conclude that the H1
data prefer the KKP set at NLO.

\begin{figure}[ht]
\begin{center}
\includegraphics[width=.75\textwidth]{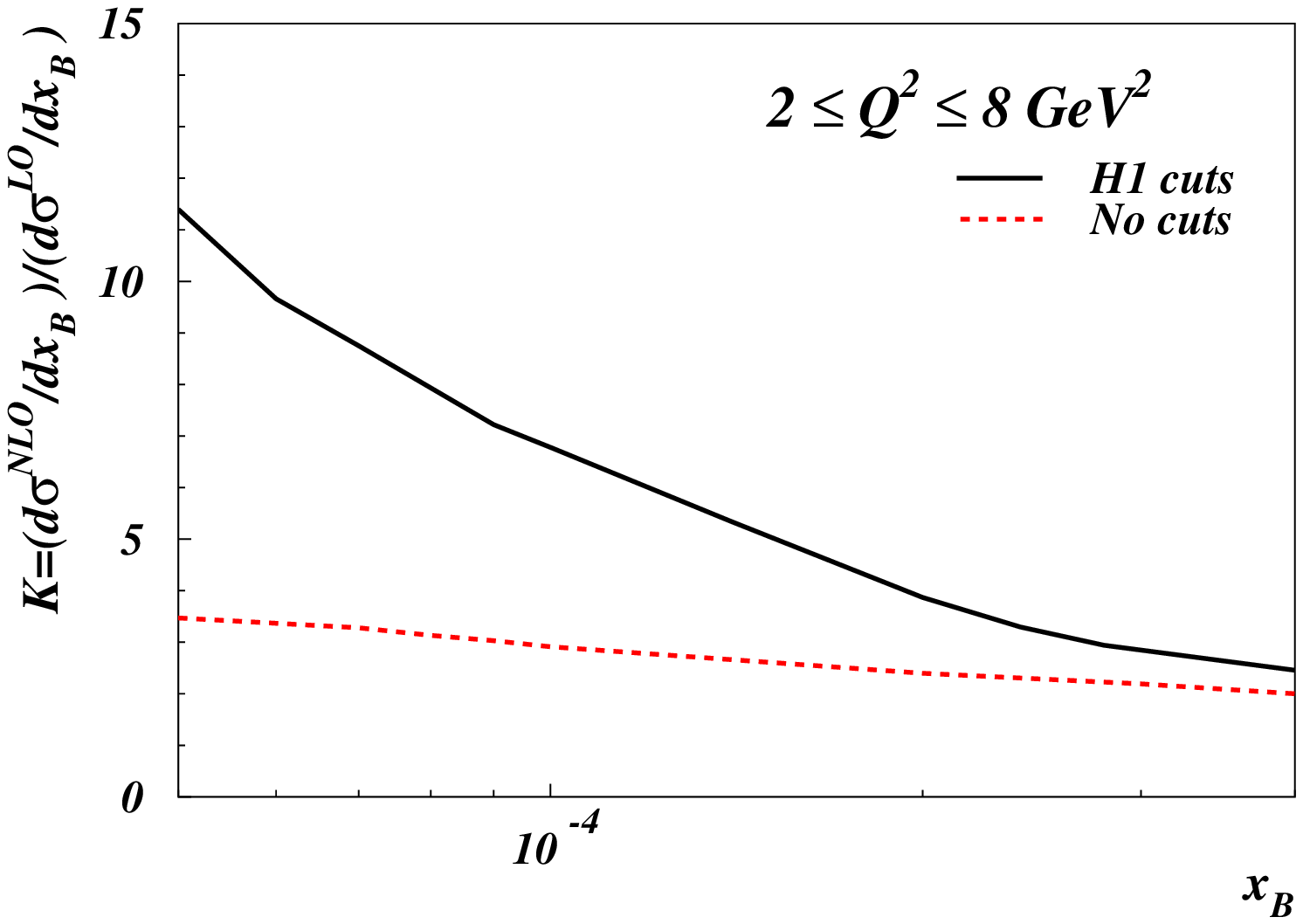}
\end{center}
\caption{$K$ factors of $d\sigma/dx_B$ for $2<Q^2<8$~GeV$^2$ and
$p_T^*>3.5$~GeV with and without the H1 forward-selection
cuts\protect\cite{Aktas:2004rb} (taken from
Ref.~\protect\refcite{Daleo:2004pn}).}
\label{fig:cuts}
\end{figure}
From the downmost frames in Figs.~\ref{fig:xs}(a)--(d), we observe that the
$K$ factors are rather sizeable, although the $\mu$ values are reasonably
large.
In Fig.~\ref{fig:cuts},\cite{Daleo:2004pn} the impact of the H1
forward-selection cuts on the $K$ factor is studied for the case of
$d\sigma/dx_B$ for $2<Q^2<8$~GeV$^2$ and $p_T^*>3.5$~GeV.
Towards the lower end of the considered $x_B$ range, the $K$ factor reaches one
order of magnitude if these cuts are imposed [see also Fig.~\ref{fig:xs}(c)].
However, if the latter are removed, the $K$ factor collapses to acceptable
values of around 3.
From this finding, we conclude that these cuts almost quench the LO cross
section.
In other words, in the extreme forward regime, the latter is effectively
generated by the $2\to3$ partonic subprocesses of Eq.~(\ref{eq:nlo}).

\begin{figure}[ht]
\begin{center}
\includegraphics[width=.75\textwidth]{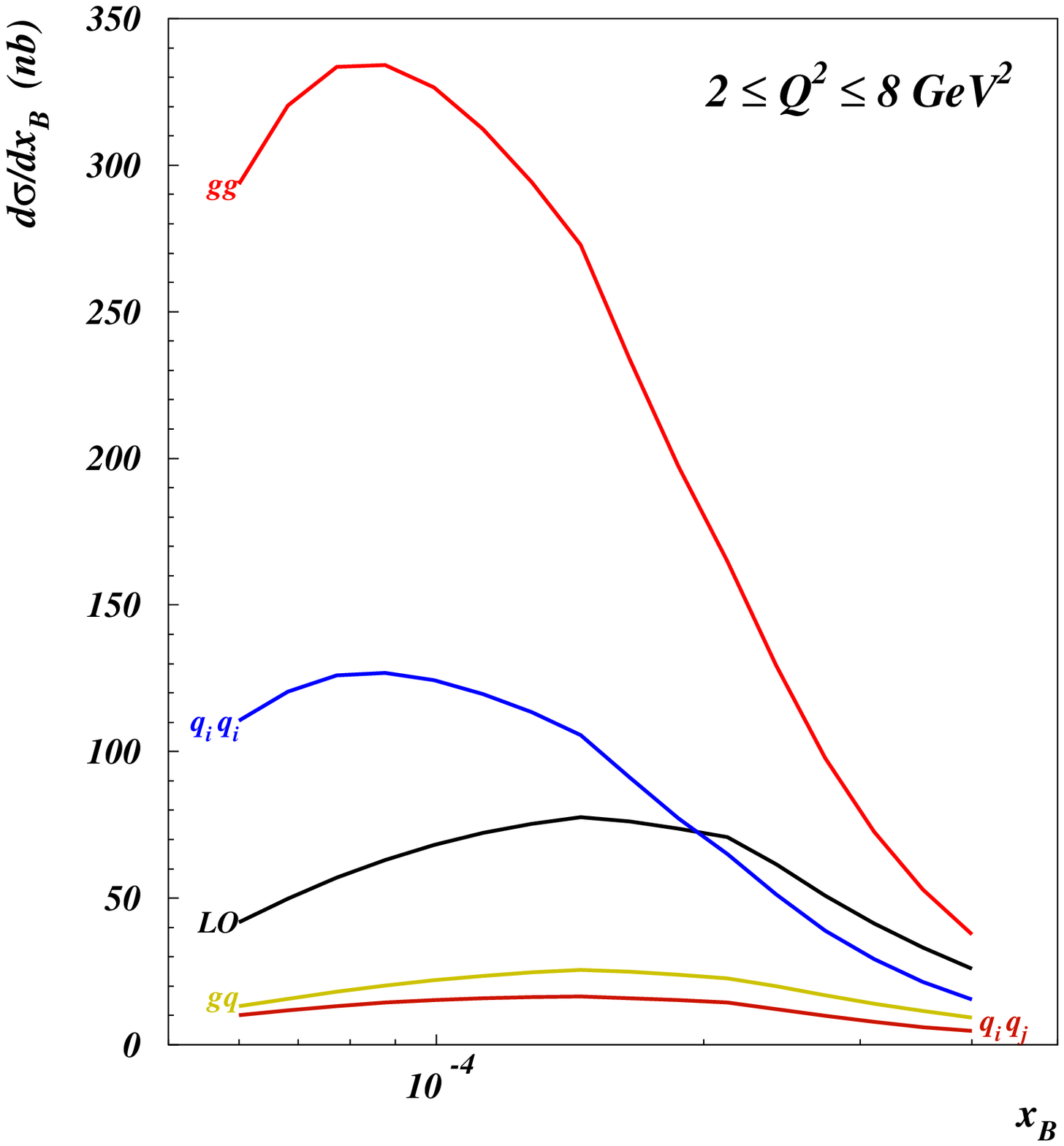}
\end{center}
\caption{Total LO contribution and NLO contributions from the four most
important $ab$ channels, where $a$ and $b$ are the partons connected with the
PDFs and FFs, respectively, to $d\sigma/dx_B$ for $2<Q^2<8$~GeV$^2$ and
$p_T^*>3.5$~GeV with the H1 forward-selection cuts\protect\cite{Aktas:2004rb}
(taken from Ref.~\protect\refcite{Daleo:2004pn}).}
\label{fig:channels}
\end{figure}
It is interesting to investigate the relative importance of the {\it tagged}
partons, i.e.\ the one ($a$) that originates from the proton and the one ($b$)
that fragments into the hadron.
In Fig.~\ref{fig:channels},\cite{Daleo:2004pn} the NLO contributions from the
four most important $ab$ channels to $d\sigma/dx_B$ for $2<Q^2<8$~GeV$^2$ and
$p_T^*>3.5$~GeV with the H1 forward-selection cuts are shown together with the
total LO contribution.
We observe that the $gg$ channel makes up approximately two thirds of the
cross section in the low-$x_B$ regime.

\section{Conclusions}
\label{sec:four}

We calculated the cross section of $ep\to e\pi^0+X$ in DIS for finite values
of $p_T^*$ at LO and NLO in the parton model of QCD\cite{kkm} using the new
AKK FFs\cite{akk} and compared it with a precise measurement by the H1
Collaboration at HERA.\cite{Adloff:1999zx,Aktas:2004rb}

We found that our LO predictions always significantly fell short of the H1
data and often exhibited deviating shapes.
However, the situation dramatically improved as we proceeded to NLO, where our
default predictions, endowed with theoretical uncertainties estimated by
moderate unphysical-scale variations, led to a satisfactory description of the
H1 data in the preponderant part of the accessed phase space.
In other words, we encountered $K$ factors much in excess of unity, except
towards the regime of asymptotic freedom characterized by large values of
$p_T^*$ and/or $Q^2$.
This was unavoidably accompanied by considerable theoretical uncertainties.
Both features suggest that a reliable interpretation of the H1 data within the
QCD-improved parton model ultimately necessitates a full
next-to-next-to-leading-order analysis, which is presently out of reach,
however.
For the time being, we conclude that the successful comparison of the H1 data
with our NLO predictions provides a useful test of the universality and the
scaling violations of the FFs, which are guaranteed by the factorization
theorem and are ruled by the DGLAP evolution equations, respectively.

Significant deviations between the H1 data and our NLO predictions only
occurred in certain corners of phase space, namely in the photoproduction
limit $Q^2\to0$, where resolved virtual photons are expected to contribute,
and in the limit $\eta\to\infty$ of the pseudorapidity
$\eta=-\ln[\tan(\theta/2)]$, where fracture functions are supposed to enter
the stage.
Both refinements were not included in our analysis.
Interestingly, distinctive deviations could not be observed towards the lowest
$x_B$ values probed, which indicates that the realm of BFKL dynamics has not
actually been accessed yet.

\section*{Acknowledgments}

The author thanks G. Kramer and M. Maniatis for their collaboration.
This work was supported in part by BMBF Grant No.\ 05~HT1GUA/4.

\end{document}